\documentclass[aps,pra,reprint,superscriptaddress]{revtex4-1}

\usepackage{graphicx}
\usepackage{siunitx} 
\usepackage{hyperref}
\hypersetup{
    colorlinks=true,
    linkcolor=black,
    citecolor=black,
    filecolor=black,
    urlcolor=black,
}
\DeclareSIUnit[number-unit-product = \ ] \dBm{dBm}
\DeclareSIUnit[number-unit-product = \ ] \nm{\nano\meter}

\begin{document}

\title{Power-efficient production of photon pairs in a tapered chalcogenide microwire}
\author{Evan Meyer-Scott}
\email{emeyersc@uwaterloo.ca}
\affiliation{Institute for Quantum Computing and Department of Physics and Astronomy, University of Waterloo, 200 University Ave W, Waterloo, Ontario, Canada N2L 3G1}
\author{Audrey Dot}
\affiliation{Institute for Quantum Computing and Department of Physics and Astronomy, University of Waterloo, 200 University Ave W, Waterloo, Ontario, Canada N2L 3G1}
\author{Raja Ahmad}
\affiliation{Department of Electrical and Computer Engineering, McGill University, 3480 University Street, Montr\'eal, Qu\'ebec, Canada H3A 2A7}
\author{Lizhu Li}
\affiliation{Department of Electrical and Computer Engineering, McGill University, 3480 University Street, Montr\'eal, Qu\'ebec, Canada H3A 2A7}
\author{Martin Rochette}
\affiliation{Department of Electrical and Computer Engineering, McGill University, 3480 University Street, Montr\'eal, Qu\'ebec, Canada H3A 2A7}
\author{Thomas Jennewein}
\affiliation{Institute for Quantum Computing and Department of Physics and Astronomy, University of Waterloo, 200 University Ave W, Waterloo, Ontario, Canada N2L 3G1}
\affiliation{Quantum Information Science Program, Canadian Institute for Advanced
Research, Toronto, ON, Canada}

\begin{abstract} Using tapered fibers of As$_2$Se$_3$ chalcogenide glass, we produce photon pairs at telecommunication wavelengths with low pump powers. We found maximum coincidences-to-accidentals ratios of $2.13\pm0.07$ for degenerate pumping with \SI{3.2}{\micro\watt} average power, and $1.33\pm0.03$ for non-degenerate pumping with \SI{1.0}{\micro\watt}  and \SI{1.5}{\micro\watt} average power of the two pumps. Our results show that the ultrahigh nonlinearity in these microwires could allow single-photon pumping to produce photon pairs, enabling the production of large entangled states, heralding of single photons after lossy transmission, and photonic quantum information processing with nonlinear optics.\end{abstract}

\maketitle

Photon pair sources are the simplest and most abundant devices used to produce quantum entanglement~\cite{RevModPhys.81.865}, and are normally based on nonlinear optical effects that require large input light intensities. With highly efficient photon sources, it becomes possible to use spontaneous parametric down-conversion (SPDC) or four-wave mixing (FWM) acting on {\em single} photons as pumps in later stages of quantum information experiments~\cite{PhysRevLett.113.173601, Hamel2014Direct-g}, enabling advanced protocols like loophole-free Bell tests via photon heralding~\cite{PhysRevX.2.021010} and photonic quantum computing~\cite{Langford2011Efficien}.

However, the conversion efficiency of one pump photon into a pair through SPDC in $\chi^{(2)}$ media has not substantially increased~\cite{Tanzilli4Highly-e,Zhang2007Correlat} beyond $10^{-6}$. Focus has shifted to developing efficient FWM in $\chi^{(3)}$ fibers and waveguides, using materials with large nonlinearities such as silicon~\cite{Takesue:2012aa} and chalcogenide glasses~\cite{Xiong:10,:/content/aip/journal/apl/98/5/10.1063/1.3549744}, and also using resonator-enhanced processes~\cite{Chen:11, Reimer2:14}, atomic vapors~\cite{PhysRevA.89.023839}, and microstructured silica fibers~\cite{Langford2011Efficien}. These would allow non-degenerate pumping with a single photon and strong pump to produce pairs.

Unfortunately, none of these sources is totally suitable for converting single photons to pairs, and it remains a challenge to design the device that offers the highest efficiency with convenient operation and low noise. Silica devices cannot reach sufficiently high efficiency due to low nonlinearity~\cite{Foster:08}, and though silicon exhibits much lower noise~\cite{Harada:2010aa}, it suffers from two-photon and free-carrier absorption~\cite{Eggleton2011Chalcoge}, limiting the maximum useable pump power. Enhancing the nonlinearity through resonators or using atomic vapors requires the input photon to be narrowband, limiting the types of initial sources for this photon.

Here we demonstrate production of photon pairs in tapered As$_2$Se$_3$ microwires, capable in principle of converting a broadband single photon into a pair with $10^{-3}$ probability~\cite{PhysRevA.90.043808}. First we characterize the four-wave mixing properties of the microwires, then we produce photon pairs from degenerate pumping, and finally produce pairs from low-power non-degenerate pumps, as would be required for a single-photon-pumping experiment. 

Our microwires combine a high nonlinear coefficient~\cite{Slusher:04} ($n_2=$ \SI{1.1e-17}{\meter\squared/\watt}) with low cross-sectional area (\SI{0.24}{\micro\meter\squared}) to produce a large waveguide nonlinear parameter ($\gamma=$ \SI{188}{\per\watt\per\meter}), $\sim$\num{100000} times larger than standard silica fibers, and ten times larger than As$_2$S$_3$ chalcogenide waveguides~\cite{:/content/aip/journal/apl/98/5/10.1063/1.3549744}. Furthermore, in contrast to on-chip waveguides, microwires are drop-in compatible with existing single-mode silica fiber, are made with lengths up to tens of centimeters, and demonstrate low loss. Coupling to the microwire is accomplished by gluing standard single-mode fiber (SMF) to the chalcogenide step-index fibers on either side of the tapered region. A coating of poly(methyl methacrylate) (PMMA) increases the mechanical robustness and influences the phasematching properties of the microwires. Broadband phasematching of up to \SI{190}{\nm} bandwidth at telecommunication wavelengths has been shown using classical four-wave mixing~\cite{Ahmad:12}. 

\begin{figure}[htp]
\includegraphics[width = \columnwidth]{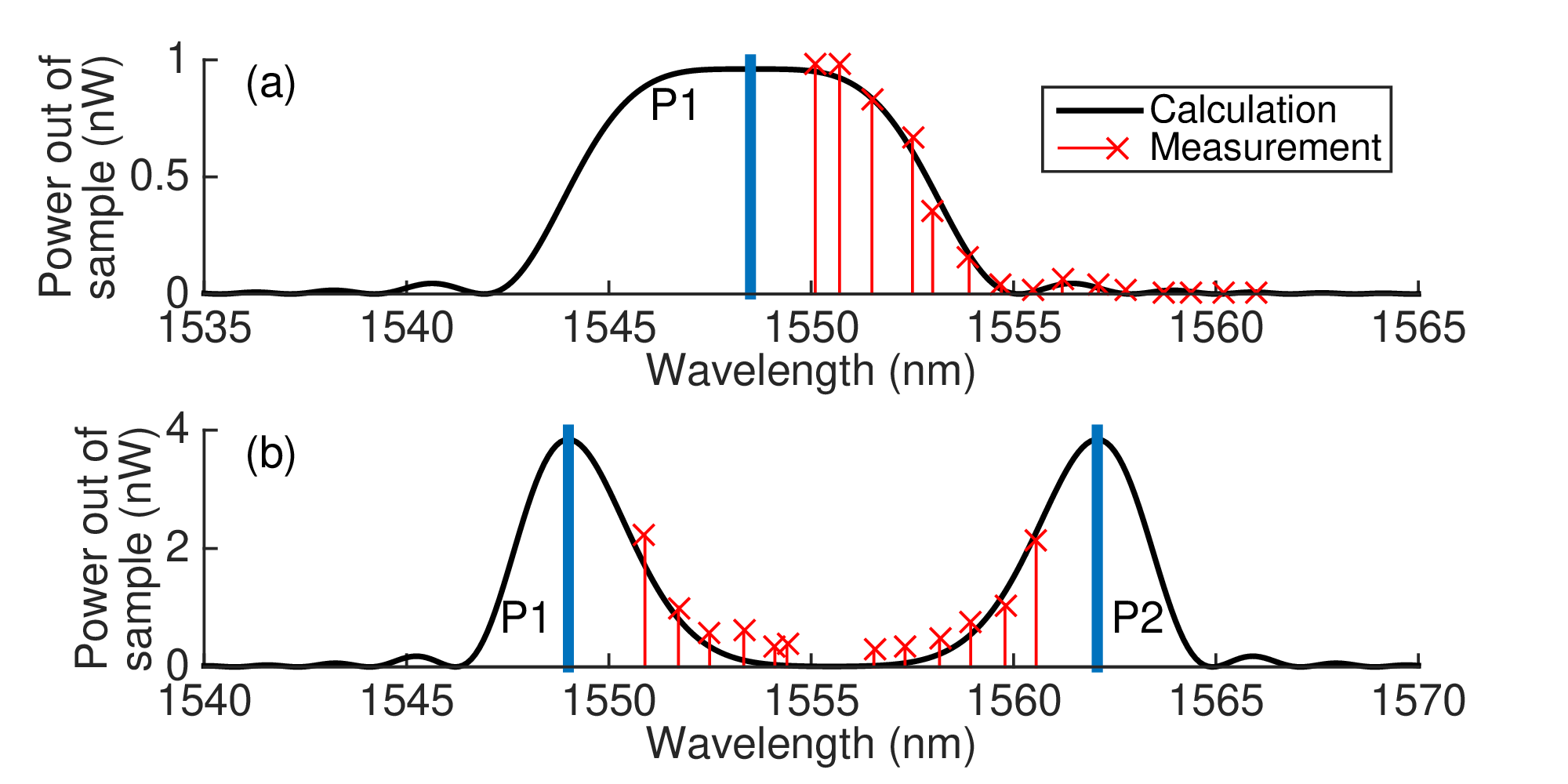}
\caption{(Color online) Phasematching obtained in the chalcogenide microwire (red markers) for (a) degenerate pumping at \SI{1548.5}{\nm} and (b) non-degenerate pumping at \SI{1549.0}{\nm} and \SI{1562.1}{\nm}. The pumps P1 and P2 are marked with thick blue bars, and the phasematching calculated directly from sample parameters is the black lines. \label{fig.phasematch}}
\end{figure}

The tapered fiber used in this work has a \SI{12}{\centi\meter} long microwire with a diameter of \SI{550}{\nano\meter}, with total insertion loss including pigtails of \SI{10}{\dB}. We estimate the breakdown of the losses as follows: inside the microwire, \SI{5}{\dB} due to sidewall roughness, absorption of the evanescent field by the polymer coating, and mode conversion in the tapering sections; at the SMF-chalcogenide interfaces, \SI{0.5}{\dB} per interface due to Fresnel reflection, and \SI{2}{\dB} per interface due to mode mismatch because the glued pigtails have been disrupted by transport. Losses per SMF-chalcogenide interface as low as 0.66 dB have been reported~\cite{Baker2012High-Non}.

Figure~\ref{fig.phasematch} shows the phasematching profile of our microwire at $\sim$\SI{1550}{\nm}, far from any zero dispersion wavelength, for a degenerate pump (a) and for two non-degenerate pumps \SI{13}{\nm} apart (b). These data were obtained by pumping and seeding with continuous-wave lasers with \SI{190}{\micro\watt} coupled power (inferred power inside the microwire after accounting for coupling losses) each. For the degenerate case, the seed laser was scanned to the shorter wavelength side of the pump, and for the non-degenerate case, the seed was scanned between the two pump wavelengths, leaving a gap in the middle where the seed and signal cross over. In both cases the output signal was filtered through a dense-wave division multiplexer (DWDM) and measured on a power meter. The values reported here have the DWDM losses factored out.

We calculated the expected phasematching and power outputs from a standard four-wave mixing treatment~\cite{tagkey2013i}, with the length, diameter, insertion loss, and nonlinear parameter as mentioned above, and calculated loss in the microwire~\cite{Baker:10} of \SI{5.1}{\decibel/\meter}. In order to find the propagation constant and effective refractive index inside the microwire, we solved the characteristic equation for a step-index fiber with As$_2$Se$_3$ core~\cite{Slusher:04} and PMMA cladding~\cite{Kasarova20071481}. Even without fitting parameters, the phasematching calculations in Fig.~\ref{fig.phasematch} agree with the measured data.

\begin{figure}
\includegraphics[width = \columnwidth]{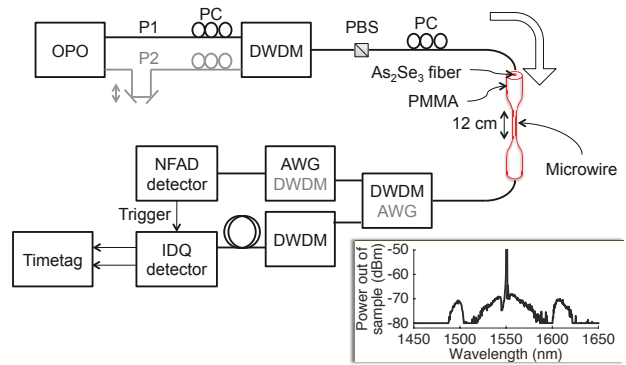}
\caption{(Color online) Experimental setup for the generation of photon pairs with a chalcogenide microwire pumped by one (P1, black) or two (P1 and P2, grey) beams from an optical parametric oscillator (OPO). For non-degenerate pumping, the two pumps were made to be copolarized with polarization controllers (PC) and a fiber polarization beamsplitter (PBS). In both degenerate and non-degenerate cases a final PC sets the polarization entering the microwire. The pump and signal/idler beams were filtered by dense wave-division multiplexers (DWDM) and an arrayed waveguide grating (AWG). The signal photon was detected by a free-running negative-feedback avalanche photodiode (NFAD) which gated the idler's single-photon detector (IDQ). Both signal and idler detection signals were recorded and timestamped by the timetag unit from Universal Quantum Devices, Inc. Inset: Raman scattering noise spectrum of the microwire pumped at a wavelength of \SI{1550}{\nm}. Two minima occur at \SI{+-40}{\nm} from the pump~\cite{Slusher:04}.  This spectrum includes a \SI{10}{\nm}-wide notch filter with \SI{30}{\decibel} blocking centered at \SI{1550}{\nm}. The pump at \SI{1550}{\nm} extends off the top of the graph to \SI{-34}{\dBm}, and our spectrometer's sensitivity is limited to \SI{-80}{\dBm}.\label{fig.scheme}}
\end{figure}

Next we produced photon pairs using a degenerate pump, with the setup shown in Fig.~\ref{fig.scheme}. The output of an optical parametric oscillator at \SI{1553.33}{\nm} with $\sim$\SI{4}{\pico\second} pulse length and \SI{76}{\mega\hertz} repetition rate was filtered through a DWDM and sent through the tapered chalcogenide microwire. The FWM output signal and idler photons were split into the \SI{1550.12}{\nm} and \SI{1556.56}{\nm} channels of a DWDM, and subsequently filtered in an AWG and DWDM respectively, giving total pump isolation of \SI{118}{\decibel} and \SI{122}{\decibel}. The signal photon was detected with a free-running InGaAs negative-feedback avalanche photodiode (NFAD) with \SI{10}{\%} detection efficiency and \num{100} dark counts per second~\cite{:/content/aip/journal/rsi/83/7/10.1063/1.4732813}, which gated an id201 InGaAs single photon detector from IDQ for the idler photon, with gate width \SI{50}{\nano\second} and \SI{20}{\%} efficiency. The gate out and detector channels from the id201 were recorded in a time-tagger to either produce timing histograms between signal and idler, or to filter with a \SI{2}{\nano\second} timing window to record coincident counts.

\begin{figure}
\includegraphics[width = \columnwidth]{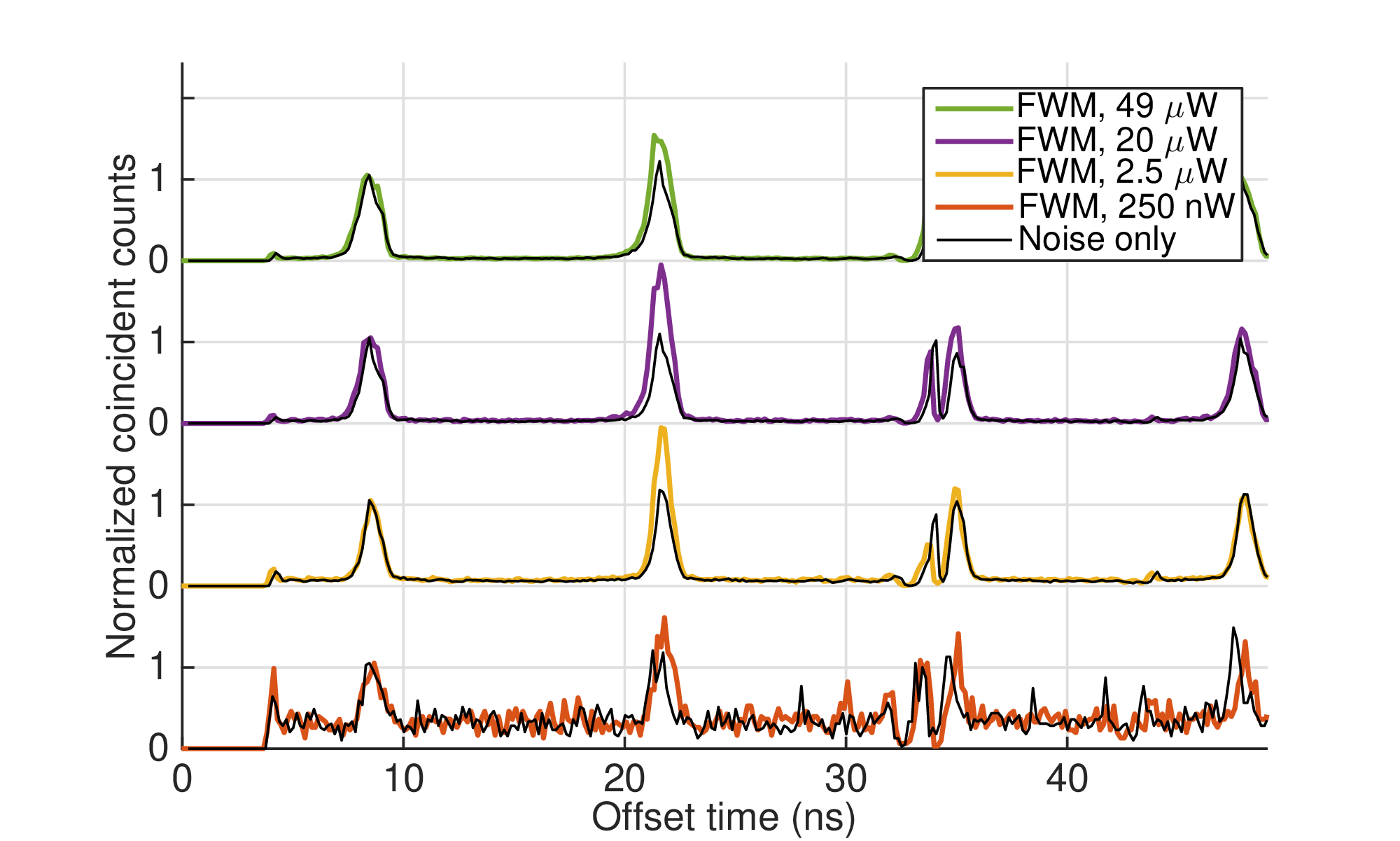}
\caption{(Color online) Timing histograms for photon pair production for coupled average pump powers of \SI{250}{\nano\watt}, \SI{2.5}{\micro\watt}, \SI{20}{\micro\watt}, and \SI{49}{\micro\watt}. The photon pairs appear in the peak at the time offset between signal and idler detection of \SI{22}{\nano\second}, which is higher than the accidental coincidence peaks at other delays. In all cases the accidentals peak height is normalized to 1. For the ``noise only'' measurement (black lines), the filter channels connected to the detectors did not conserve energy with the pump, leading to the disappearance of the legitimate photon pairs. The double peak at \SI{35}{\nano\second} could be due to the timing electronics in the IDQ detector or timetagger, but no events are lost.\label{fig.timing}}
\end{figure}

The timing histograms in Fig.~\ref{fig.timing} show the presence of photon pairs, as evidenced by a higher peak at the time delay of \SI{22}{\nano\second} than the background peaks at other time delays. At the lowest power, detector dark counts at random times become prominent, while at the highest power, accidental coincidences from multiple photon pairs make a large contribution to the noise. When the idler channel is moved to \SI{1558.17}{\nm} (black lines in Fig.~\ref{fig.timing}) such that photon pairs detected would not conserve energy and could not be from four-wave mixing, the peak at \SI{22}{\nano\second} falls to the same height as the others.

\begin{figure}
\includegraphics[width = \columnwidth]{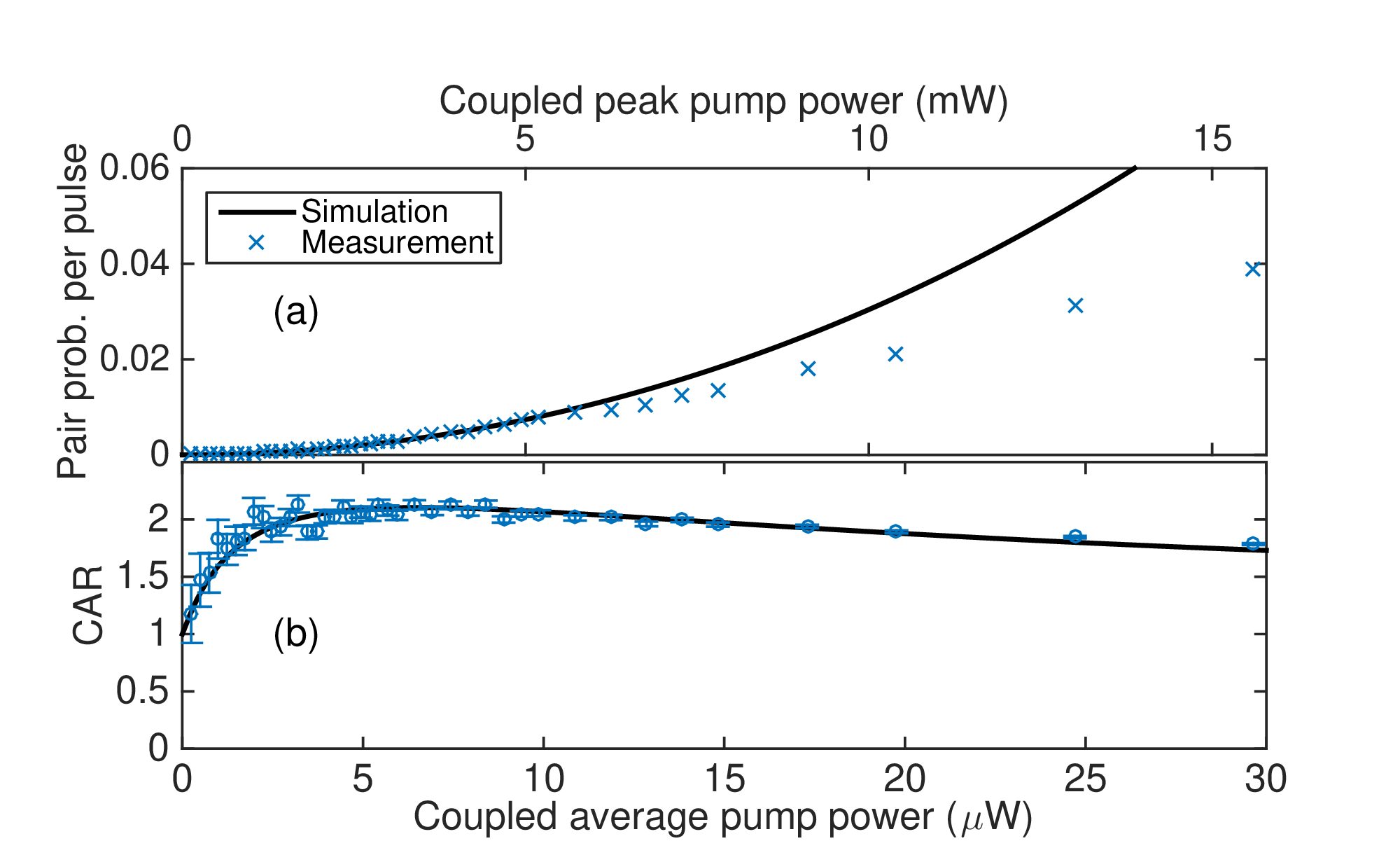}
\caption{(Color online) Pair probability per laser pulse with accidentals subtracted (a) and coincidences-to-accidentals ratio (b) as measured for degenerate-pumping FWM in our chalcogenide microwire. The curves are from a FWM simulation including measured background counts, with the pump pulse length inside the microwire as a fitting parameter. $CAR >1$ indicates photon pairs are detected above the noise. Error bars are based on poissonian uncertainty in photon counts, and are smaller than symbol size for (a). The x-axes on figures (a) and (b) coincide and show both peak and average power.\label{fig.CAR}}
\end{figure}

As seen in Fig.~\ref{fig.CAR}(a), the number of pairs per pulse increases quadratically with pump power, with dropoff seen at high count rates due to the dead time of the NFAD detector. To compare with previous photon pair sources, we calculate the number of pairs produced per second, per nanometer of signal bandwidth, per milliwatt of average pump power as \SI{2.5e8}{pairs\per\second\per\nm\per\milli\watt} for the data point with \SI{30}{\micro\watt} pump power. This is significantly above what is possible in $\chi^{(2)}$ crystals~\cite{Steinlechner:13}, and rivals the highest values reported in silicon~\cite{Engin:13}. Note that our filters had \SI{0.5}{\nm} bandwidth, and milliwatt pump powers are too large to avoid damage to the microwire.

In Fig.~\ref{fig.CAR}(b), the coincidences-to-accidentals ratio reaches its maximum value of $2.13\pm0.07$ at a coupled pump power of \SI{3.2}{\micro\watt}. This ratio is defined as $CAR=C/A$, where $C$ is the total number of coincident counts in the main coincidence peak and $A$ is the number of accidental coincidences, and has a lower bound of $CAR=1$ for no timing correlation. At a coupled pump power of \SI{490}{\nano\watt}, the CAR was $1.5\pm0.2$, and statistical significance increased with increasing pump power. Here the coincidences are collected at the \SI{22}{\nano\second} offset time of Fig.~\ref{fig.timing}, while the accidentals, which are due to detector dark counts, double-pair emissions from FWM, and Raman and other optical noise, are collected at \SI{9}{\nano\second}, which allows an estimation of the contribution of accidentals to the main coincidence peak. At low power, the CAR decreases due to the small number of real photon pairs compared to noise photons, and at high power, the CAR slowly decreases due to double-pair emissions from four-wave mixing. The fits to pair probability and CAR come from the FWM calculation described above, which feeds into a quantum-optical simulation including FWM, loss, detector models, and background counts measured with a continuous-wave pump. Here the pump pulse length inside the microwire is used as a fitting parameter and found to be \SI{25}{\pico\second}. 

In order to approach the regime of converting a single photon into a pair, the two pump photons absorbed in the FWM process must be non-degenerate in wavelength. Since FWM probability goes as the product of the pump peak powers, it is important that both pumps be synchronously pulsed, rather than continuous-wave. To demonstrate non-degenerate pumping, we took two outputs from an optical parametric oscillator, passing one through a trombone delay line to synchronize the pulses in the microwire as in the grey lines and text in Fig.~\ref{fig.scheme}.

\begin{figure}
\includegraphics[width = \columnwidth]{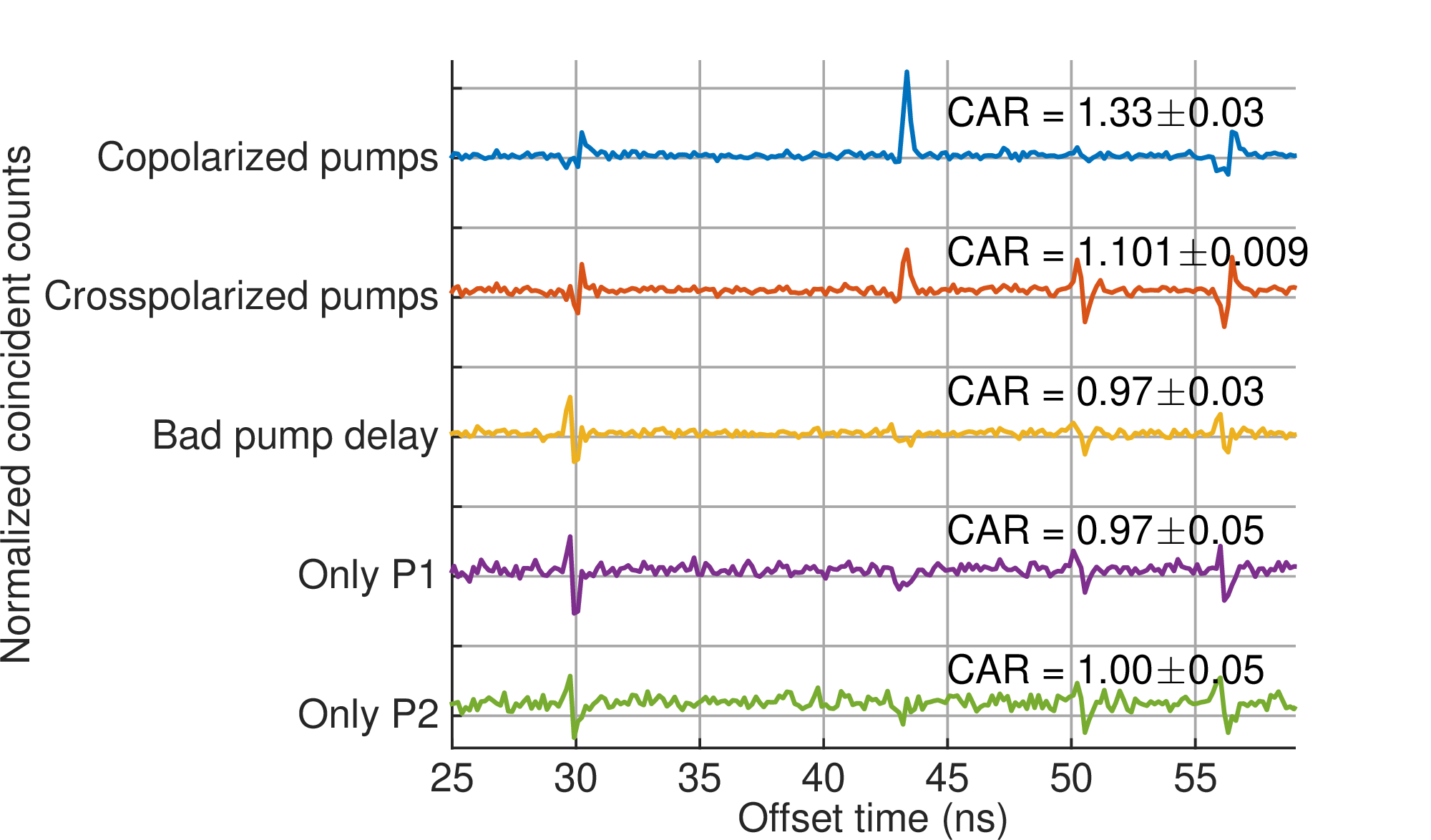}
\caption{(Color online) Timing histograms for photon pair production for non-degenerate pumping. The photon pairs appear in the peak at timing delay \SI{43}{\nano\second} only for the top two cases, with  copolarized and crosspolarized pumps respectively, giving $CAR>1$. In the other cases, where FWM is not expected, the peak at \SI{43}{\nano\second} vanishes and $CAR=1$ within error. Here each dataset has had the background ``noise only'' counts subtracted.\label{fig.timingnon}} 
\end{figure}

To find evidence of photon pairs with non-degenerate pumping, we took timing histograms as shown in Fig.~\ref{fig.timingnon}. We present a number of different scenarios: co- and cross-polarized pumps, changing the delay between the two pumps so they do not overlap in the microwire, and removing either of the pumps. Only the first two cases, with the crosspolarized pumps producing 4/9 the number of pairs of the copolarized pumps~\cite{tagkey2013i}, provide a  peak at the proper delay after background subtraction, indicating these photons are due to non-degenerate FWM as desired. The maximum CAR for non-degenerate pumping was $1.33\pm0.03$ with \SI{1.0}{\micro\watt} and \SI{1.5}{\micro\watt} coupled powers of the two pumps, while the $CAR$ for the lowest asymmetric pump power was $1.17\pm0.06$ with coupled pump powers of \SI{480}{\nano\watt} and \SI{1.5}{\micro\watt}.

\begin{figure}
\includegraphics[width = \columnwidth]{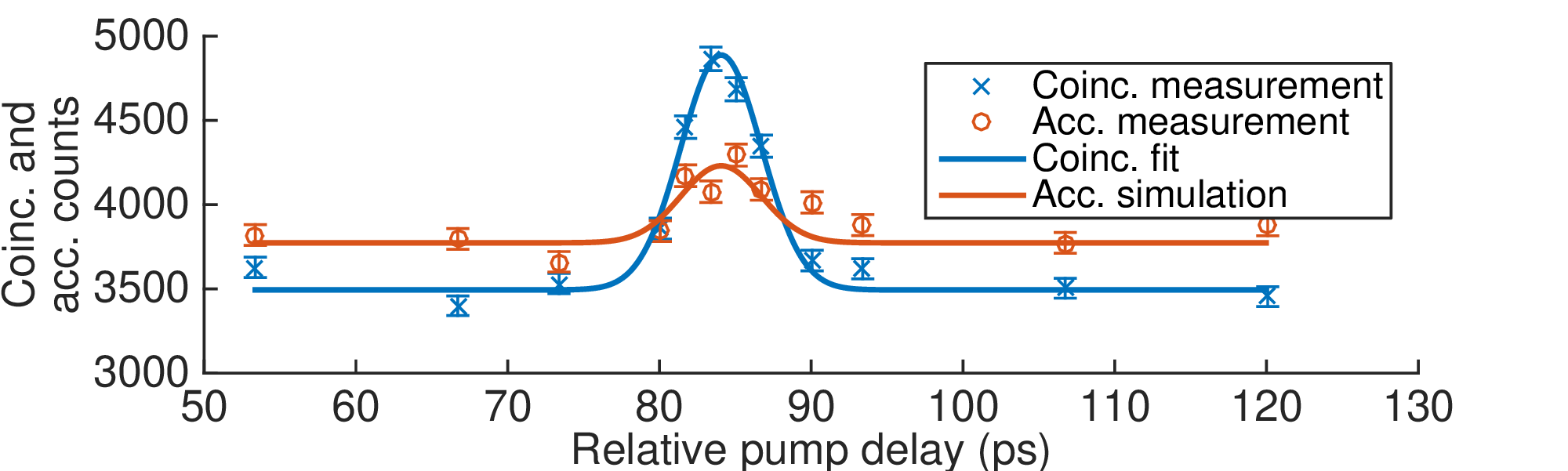}
\caption{(Color online) Coincident and accidental counts versus time delay between pump pulses for non-degenerate pumping. Photon pairs are only produced when the pumps overlap in the microwire; accidentals measured \SI{13}{\nano\second} (one pump period) later also increase as the pumps overlap due to multi-pair emissions. Here the accidentals are higher than the coincidences outside the peak because the accidentals were collected from a more efficient part of the IDQ detector's gate. The line on the coincidence data is a gaussian fit which is fed into a FWM simulation to generate the accidentals curve.\label{fig.delay}}
\end{figure}

Unfortunately due to bad phasematching (see Fig.~\ref{fig.phasematch}) and the extra noise brought by having two pumps, these data are not as clear as the degenerate case. A more convincing measurement is one of coincidences and accidentals versus time delay between the two pumps, where photon pairs are produced only when the two pumps overlap, on top a constant background caused by spontaneous Raman scattering. In Fig.~\ref{fig.delay}, we show these data for P1~ =~\SI{1551.72}{\nm}, P2~=~\SI{1561.42}{\nm}, and the signal and idler wavelengths \SI{1554.13}{\nm} and \SI{1558.98}{\nm}.
The coincidence curve is a gaussian fit to the data, which is fed into a quantum-optical FWM simulation that includes higher-order emission but no other nonlinearities to find the expected number of accidentals. The simulated accidentals agree with the data, indicating that the increased accidentals when the pumps overlap are due only to photon emission statistics of the FWM process.

To achieve in future the splitting of a single photon into two will require a single-photon pump of \SI{10}{\pico\watt} average power. Due to the narrow phasematching shown in Fig.~\ref{fig.phasematch}, we are currently forced to collect photon pairs in a wavelength region close to the pumps that is contaminated with noise (inset in Fig.~\ref{fig.scheme}). The lowest pump power for non-degenerate pumping with which we found $CAR>1$ was \SI{480}{\nano\watt}. Additionally, reaching $10^{-3}$ conversion efficiency~\cite{PhysRevA.90.043808} would require an average power for the other pump of \SI{130}{\micro\watt}, whereas we found $CAR>1$ for non-degenerate pump power only up to \SI{3}{\micro\watt}. This does not mean that the high conversion efficiency cannot be achieved, but it requires moving the signal and idler much farther in wavelength from the strong pump to avoid noise and broadening their phasematching bandwidths to increase efficiency. By carefully controlling core size and cladding material, it will be possible in future to fabricate microwires with engineered dispersion to reproduce broadband phasematching previously obtained~\cite{Ahmad:12}. If the current \SI{0.5}{\nm} photon bandwidth can be expanded to \SI{40}{\nm}, centered \SI{100}{\nm} from the pump, \num{80} times more efficient conversion is expected. These photon pairs would be produced in the Raman gain dips or even beyond the gain peaks, which would improve CAR drastically as shown in chalcogenide waveguides~\cite{:/content/aip/journal/jap/112/12/10.1063/1.4769740,Collins:12}.

We have presented evidence of photon pairs produced in an As$_2$Se$_3$ chalcogenide microwire for both degenerate and non-degenerate pumping. Because this device is a tapered fibre, the coupling of optical signals in and out is straightforward and stable with room for improvement in efficiency, making the system very interesting for future applications. Through timing analysis of photon pairs, we found that the coincidences-to-accidentals ratio maintains a value of $CAR>1$ over a wide range of pump powers, and inferred a maximum pair production rate inside the microwire (limited by detector dead time) of \SI{2.9e6}{pairs/s}. We look forward to reducing the background noise by engineering the phasematching conditions and pushing closer to $10^{-3}$ efficiency for converting a single photon into a pair.

We acknowledge support from NSERC, Ontario Ministry of Research and Innovation, CIFAR, FedDev Ontario, Industry Canada, and CFI. We are grateful to Coractive High-Tech for providing the chalcogenide glass used in the experiments.


\begin{thebibliography}{28}%
\makeatletter
\providecommand \@ifxundefined [1]{%
 \@ifx{#1\undefined}
}%
\providecommand \@ifnum [1]{%
 \ifnum #1\expandafter \@firstoftwo
 \else \expandafter \@secondoftwo
 \fi
}%
\providecommand \@ifx [1]{%
 \ifx #1\expandafter \@firstoftwo
 \else \expandafter \@secondoftwo
 \fi
}%
\providecommand \natexlab [1]{#1}%
\providecommand \enquote  [1]{``#1''}%
\providecommand \bibnamefont  [1]{#1}%
\providecommand \bibfnamefont [1]{#1}%
\providecommand \citenamefont [1]{#1}%
\providecommand \href@noop [0]{\@secondoftwo}%
\providecommand \href [0]{\begingroup \@sanitize@url \@href}%
\providecommand \@href[1]{\@@startlink{#1}\@@href}%
\providecommand \@@href[1]{\endgroup#1\@@endlink}%
\providecommand \@sanitize@url [0]{\catcode `\\12\catcode `\$12\catcode
  `\&12\catcode `\#12\catcode `\^12\catcode `\_12\catcode `\%12\relax}%
\providecommand \@@startlink[1]{}%
\providecommand \@@endlink[0]{}%
\providecommand \url  [0]{\begingroup\@sanitize@url \@url }%
\providecommand \@url [1]{\endgroup\@href {#1}{\urlprefix }}%
\providecommand \urlprefix  [0]{URL }%
\providecommand \Eprint [0]{\href }%
\providecommand \doibase [0]{http://dx.doi.org/}%
\providecommand \selectlanguage [0]{\@gobble}%
\providecommand \bibinfo  [0]{\@secondoftwo}%
\providecommand \bibfield  [0]{\@secondoftwo}%
\providecommand \translation [1]{[#1]}%
\providecommand \BibitemOpen [0]{}%
\providecommand \bibitemStop [0]{}%
\providecommand \bibitemNoStop [0]{.\EOS\space}%
\providecommand \EOS [0]{\spacefactor3000\relax}%
\providecommand \BibitemShut  [1]{\csname bibitem#1\endcsname}%
\let\auto@bib@innerbib\@empty
\bibitem [{\citenamefont {Horodecki}\ \emph {et~al.}(2009)\citenamefont
  {Horodecki}, \citenamefont {Horodecki}, \citenamefont {Horodecki},\ and\
  \citenamefont {Horodecki}}]{RevModPhys.81.865}%
  \BibitemOpen
  \bibfield  {author} {\bibinfo {author} {\bibfnamefont {R.}~\bibnamefont
  {Horodecki}}, \bibinfo {author} {\bibfnamefont {P.}~\bibnamefont
  {Horodecki}}, \bibinfo {author} {\bibfnamefont {M.}~\bibnamefont
  {Horodecki}}, \ and\ \bibinfo {author} {\bibfnamefont {K.}~\bibnamefont
  {Horodecki}},\ }\href {\doibase 10.1103/RevModPhys.81.865} {\bibfield
  {journal} {\bibinfo  {journal} {Rev. Mod. Phys.}\ }\textbf {\bibinfo {volume}
  {81}},\ \bibinfo {pages} {865} (\bibinfo {year} {2009})}\BibitemShut
  {NoStop}%
\bibitem [{\citenamefont {Guerreiro}\ \emph {et~al.}(2014)\citenamefont
  {Guerreiro}, \citenamefont {Martin}, \citenamefont {Sanguinetti},
  \citenamefont {Pelc}, \citenamefont {Langrock}, \citenamefont {Fejer},
  \citenamefont {Gisin}, \citenamefont {Zbinden}, \citenamefont {Sangouard},\
  and\ \citenamefont {Thew}}]{PhysRevLett.113.173601}%
  \BibitemOpen
  \bibfield  {author} {\bibinfo {author} {\bibfnamefont {T.}~\bibnamefont
  {Guerreiro}}, \bibinfo {author} {\bibfnamefont {A.}~\bibnamefont {Martin}},
  \bibinfo {author} {\bibfnamefont {B.}~\bibnamefont {Sanguinetti}}, \bibinfo
  {author} {\bibfnamefont {J.~S.}\ \bibnamefont {Pelc}}, \bibinfo {author}
  {\bibfnamefont {C.}~\bibnamefont {Langrock}}, \bibinfo {author}
  {\bibfnamefont {M.~M.}\ \bibnamefont {Fejer}}, \bibinfo {author}
  {\bibfnamefont {N.}~\bibnamefont {Gisin}}, \bibinfo {author} {\bibfnamefont
  {H.}~\bibnamefont {Zbinden}}, \bibinfo {author} {\bibfnamefont
  {N.}~\bibnamefont {Sangouard}}, \ and\ \bibinfo {author} {\bibfnamefont
  {R.~T.}\ \bibnamefont {Thew}},\ }\href {\doibase
  10.1103/PhysRevLett.113.173601} {\bibfield  {journal} {\bibinfo  {journal}
  {Phys. Rev. Lett.}\ }\textbf {\bibinfo {volume} {113}},\ \bibinfo {pages}
  {173601} (\bibinfo {year} {2014})}\BibitemShut {NoStop}%
\bibitem [{\citenamefont {Hamel}\ \emph {et~al.}(2014)\citenamefont {Hamel},
  \citenamefont {Shalm}, \citenamefont {H{\"u}bel}, \citenamefont {Miller},
  \citenamefont {Marsili}, \citenamefont {Verma}, \citenamefont {Mirin},
  \citenamefont {Nam}, \citenamefont {Resch},\ and\ \citenamefont
  {Jennewein}}]{Hamel2014Direct-g}%
  \BibitemOpen
  \bibfield  {author} {\bibinfo {author} {\bibfnamefont {D.~R.}\ \bibnamefont
  {Hamel}}, \bibinfo {author} {\bibfnamefont {L.~K.}\ \bibnamefont {Shalm}},
  \bibinfo {author} {\bibfnamefont {H.}~\bibnamefont {H{\"u}bel}}, \bibinfo
  {author} {\bibfnamefont {A.~J.}\ \bibnamefont {Miller}}, \bibinfo {author}
  {\bibfnamefont {F.}~\bibnamefont {Marsili}}, \bibinfo {author} {\bibfnamefont
  {V.~B.}\ \bibnamefont {Verma}}, \bibinfo {author} {\bibfnamefont {R.~P.}\
  \bibnamefont {Mirin}}, \bibinfo {author} {\bibfnamefont {S.~W.}\ \bibnamefont
  {Nam}}, \bibinfo {author} {\bibfnamefont {K.~J.}\ \bibnamefont {Resch}}, \
  and\ \bibinfo {author} {\bibfnamefont {T.}~\bibnamefont {Jennewein}},\ }\href
  {http://dx.doi.org/10.1038/nphoton.2014.218} {\bibfield  {journal} {\bibinfo
  {journal} {Nature Photon.}\ }\textbf {\bibinfo {volume} {8}},\ \bibinfo
  {pages} {801} (\bibinfo {year} {2014})}\BibitemShut {NoStop}%
\bibitem [{\citenamefont {Cabello}\ and\ \citenamefont
  {Sciarrino}(2012)}]{PhysRevX.2.021010}%
  \BibitemOpen
  \bibfield  {author} {\bibinfo {author} {\bibfnamefont {A.}~\bibnamefont
  {Cabello}}\ and\ \bibinfo {author} {\bibfnamefont {F.}~\bibnamefont
  {Sciarrino}},\ }\href@noop {} {\bibfield  {journal} {\bibinfo  {journal}
  {Phys. Rev. X}\ }\textbf {\bibinfo {volume} {2}},\ \bibinfo {pages}
  {{021010}} (\bibinfo {year} {2012})}\BibitemShut {NoStop}%
\bibitem [{\citenamefont {Langford}\ \emph {et~al.}(2011)\citenamefont
  {Langford}, \citenamefont {Ramelow}, \citenamefont {Prevedel}, \citenamefont
  {Munro}, \citenamefont {Milburn},\ and\ \citenamefont
  {Zeilinger}}]{Langford2011Efficien}%
  \BibitemOpen
  \bibfield  {author} {\bibinfo {author} {\bibfnamefont {N.~K.}\ \bibnamefont
  {Langford}}, \bibinfo {author} {\bibfnamefont {S.}~\bibnamefont {Ramelow}},
  \bibinfo {author} {\bibfnamefont {R.}~\bibnamefont {Prevedel}}, \bibinfo
  {author} {\bibfnamefont {W.~J.}\ \bibnamefont {Munro}}, \bibinfo {author}
  {\bibfnamefont {G.~J.}\ \bibnamefont {Milburn}}, \ and\ \bibinfo {author}
  {\bibfnamefont {A.}~\bibnamefont {Zeilinger}},\ }\href
  {http://dx.doi.org/10.1038/nature10463} {\bibfield  {journal} {\bibinfo
  {journal} {Nature}\ }\textbf {\bibinfo {volume} {478}},\ \bibinfo {pages}
  {360} (\bibinfo {year} {2011})}\BibitemShut {NoStop}%
\bibitem [{\citenamefont {Tanzilli}\ \emph {et~al.}(2001)\citenamefont
  {Tanzilli}, \citenamefont {de~Riedmatten}, \citenamefont {Tittel},
  \citenamefont {Zbinden}, \citenamefont {Baldi}, \citenamefont {De~Micheli},
  \citenamefont {Ostrowsky},\ and\ \citenamefont {Gisin}}]{Tanzilli4Highly-e}%
  \BibitemOpen
  \bibfield  {author} {\bibinfo {author} {\bibfnamefont {S.}~\bibnamefont
  {Tanzilli}}, \bibinfo {author} {\bibfnamefont {H.}~\bibnamefont
  {de~Riedmatten}}, \bibinfo {author} {\bibfnamefont {H.}~\bibnamefont
  {Tittel}}, \bibinfo {author} {\bibfnamefont {H.}~\bibnamefont {Zbinden}},
  \bibinfo {author} {\bibfnamefont {P.}~\bibnamefont {Baldi}}, \bibinfo
  {author} {\bibfnamefont {M.}~\bibnamefont {De~Micheli}}, \bibinfo {author}
  {\bibfnamefont {D.~B.}\ \bibnamefont {Ostrowsky}}, \ and\ \bibinfo {author}
  {\bibfnamefont {N.}~\bibnamefont {Gisin}},\ }\href {\doibase
  10.1049/el:20010009} {\bibfield  {journal} {\bibinfo  {journal} {Electron.
  Lett.}\ }\textbf {\bibinfo {volume} {37}},\ \bibinfo {pages} {26} (\bibinfo
  {year} {2001})}\BibitemShut {NoStop}%
\bibitem [{\citenamefont {Zhang}\ \emph {et~al.}(2007)\citenamefont {Zhang},
  \citenamefont {Xie}, \citenamefont {Takesue}, \citenamefont {Nam},
  \citenamefont {Langrock}, \citenamefont {Fejer},\ and\ \citenamefont
  {Yamamoto}}]{Zhang2007Correlat}%
  \BibitemOpen
  \bibfield  {author} {\bibinfo {author} {\bibfnamefont {Q.}~\bibnamefont
  {Zhang}}, \bibinfo {author} {\bibfnamefont {X.}~\bibnamefont {Xie}}, \bibinfo
  {author} {\bibfnamefont {H.}~\bibnamefont {Takesue}}, \bibinfo {author}
  {\bibfnamefont {S.~W.}\ \bibnamefont {Nam}}, \bibinfo {author} {\bibfnamefont
  {C.}~\bibnamefont {Langrock}}, \bibinfo {author} {\bibfnamefont {M.~M.}\
  \bibnamefont {Fejer}}, \ and\ \bibinfo {author} {\bibfnamefont
  {Y.}~\bibnamefont {Yamamoto}},\ }\href
  {http://www.opticsexpress.org/abstract.cfm?URI=oe-15-16-10288} {\bibfield
  {journal} {\bibinfo  {journal} {Opt. Express}\ }\textbf {\bibinfo {volume}
  {15}},\ \bibinfo {pages} {10288} (\bibinfo {year} {2007})}\BibitemShut
  {NoStop}%
\bibitem [{\citenamefont {Takesue}(2012)}]{Takesue:2012aa}%
  \BibitemOpen
  \bibfield  {author} {\bibinfo {author} {\bibfnamefont {H.}~\bibnamefont
  {Takesue}},\ }\href@noop {} {\bibfield  {journal} {\bibinfo  {journal} {IEEE
  J. Sel. Top. Quantum Electron.}\ }\textbf {\bibinfo {volume} {18}},\ \bibinfo
  {pages} {1722} (\bibinfo {year} {2012})}\BibitemShut {NoStop}%
\bibitem [{\citenamefont {Xiong}\ \emph {et~al.}(2010)\citenamefont {Xiong},
  \citenamefont {Helt}, \citenamefont {Judge}, \citenamefont {Marshall},
  \citenamefont {Steel}, \citenamefont {Sipe},\ and\ \citenamefont
  {Eggleton}}]{Xiong:10}%
  \BibitemOpen
  \bibfield  {author} {\bibinfo {author} {\bibfnamefont {C.}~\bibnamefont
  {Xiong}}, \bibinfo {author} {\bibfnamefont {L.~G.}\ \bibnamefont {Helt}},
  \bibinfo {author} {\bibfnamefont {A.~C.}\ \bibnamefont {Judge}}, \bibinfo
  {author} {\bibfnamefont {G.~D.}\ \bibnamefont {Marshall}}, \bibinfo {author}
  {\bibfnamefont {M.~J.}\ \bibnamefont {Steel}}, \bibinfo {author}
  {\bibfnamefont {J.~E.}\ \bibnamefont {Sipe}}, \ and\ \bibinfo {author}
  {\bibfnamefont {B.~J.}\ \bibnamefont {Eggleton}},\ }\href {\doibase
  10.1364/OE.18.016206} {\bibfield  {journal} {\bibinfo  {journal} {Opt.
  Express}\ }\textbf {\bibinfo {volume} {18}},\ \bibinfo {pages} {16206}
  (\bibinfo {year} {2010})}\BibitemShut {NoStop}%
\bibitem [{\citenamefont {Xiong}\ \emph {et~al.}(2011)\citenamefont {Xiong},
  \citenamefont {Marshall}, \citenamefont {Peruzzo}, \citenamefont {Lobino},
  \citenamefont {Clark}, \citenamefont {Choi}, \citenamefont {Madden},
  \citenamefont {Natarajan}, \citenamefont {Tanner}, \citenamefont {Hadfield},
  \citenamefont {Dorenbos}, \citenamefont {Zijlstra}, \citenamefont {Zwiller},
  \citenamefont {Thompson}, \citenamefont {Rarity}, \citenamefont {Steel},
  \citenamefont {Luther-Davies}, \citenamefont {Eggleton},\ and\ \citenamefont
  {O'Brien}}]{:/content/aip/journal/apl/98/5/10.1063/1.3549744}%
  \BibitemOpen
  \bibfield  {author} {\bibinfo {author} {\bibfnamefont {C.}~\bibnamefont
  {Xiong}}, \bibinfo {author} {\bibfnamefont {G.~D.}\ \bibnamefont {Marshall}},
  \bibinfo {author} {\bibfnamefont {A.}~\bibnamefont {Peruzzo}}, \bibinfo
  {author} {\bibfnamefont {M.}~\bibnamefont {Lobino}}, \bibinfo {author}
  {\bibfnamefont {A.~S.}\ \bibnamefont {Clark}}, \bibinfo {author}
  {\bibfnamefont {D.-Y.}\ \bibnamefont {Choi}}, \bibinfo {author}
  {\bibfnamefont {S.~J.}\ \bibnamefont {Madden}}, \bibinfo {author}
  {\bibfnamefont {C.~M.}\ \bibnamefont {Natarajan}}, \bibinfo {author}
  {\bibfnamefont {M.~G.}\ \bibnamefont {Tanner}}, \bibinfo {author}
  {\bibfnamefont {R.~H.}\ \bibnamefont {Hadfield}}, \bibinfo {author}
  {\bibfnamefont {S.~N.}\ \bibnamefont {Dorenbos}}, \bibinfo {author}
  {\bibfnamefont {T.}~\bibnamefont {Zijlstra}}, \bibinfo {author}
  {\bibfnamefont {V.}~\bibnamefont {Zwiller}}, \bibinfo {author} {\bibfnamefont
  {M.~G.}\ \bibnamefont {Thompson}}, \bibinfo {author} {\bibfnamefont {J.~G.}\
  \bibnamefont {Rarity}}, \bibinfo {author} {\bibfnamefont {M.~J.}\
  \bibnamefont {Steel}}, \bibinfo {author} {\bibfnamefont {B.}~\bibnamefont
  {Luther-Davies}}, \bibinfo {author} {\bibfnamefont {B.~J.}\ \bibnamefont
  {Eggleton}}, \ and\ \bibinfo {author} {\bibfnamefont {J.~L.}\ \bibnamefont
  {O'Brien}},\ }\href {\doibase http://dx.doi.org/10.1063/1.3549744} {\bibfield
   {journal} {\bibinfo  {journal} {Appl. Phys. Lett.}\ }\textbf {\bibinfo
  {volume} {98}},\ \bibinfo {eid} {051101} (\bibinfo {year}
  {2011})}\BibitemShut {NoStop}%
\bibitem [{\citenamefont {Chen}\ \emph {et~al.}(2011)\citenamefont {Chen},
  \citenamefont {Levine}, \citenamefont {Fan},\ and\ \citenamefont
  {Migdall}}]{Chen:11}%
  \BibitemOpen
  \bibfield  {author} {\bibinfo {author} {\bibfnamefont {J.}~\bibnamefont
  {Chen}}, \bibinfo {author} {\bibfnamefont {Z.~H.}\ \bibnamefont {Levine}},
  \bibinfo {author} {\bibfnamefont {J.}~\bibnamefont {Fan}}, \ and\ \bibinfo
  {author} {\bibfnamefont {A.~L.}\ \bibnamefont {Migdall}},\ }\href {\doibase
  10.1364/OE.19.001470} {\bibfield  {journal} {\bibinfo  {journal} {Opt.
  Express}\ }\textbf {\bibinfo {volume} {19}},\ \bibinfo {pages} {1470}
  (\bibinfo {year} {2011})}\BibitemShut {NoStop}%
\bibitem [{\citenamefont {Reimer}\ \emph {et~al.}(2014)\citenamefont {Reimer},
  \citenamefont {Caspani}, \citenamefont {Clerici}, \citenamefont {Ferrera},
  \citenamefont {Kues}, \citenamefont {Peccianti}, \citenamefont {Pasquazi},
  \citenamefont {Razzari}, \citenamefont {Little}, \citenamefont {Chu},
  \citenamefont {Moss},\ and\ \citenamefont {Morandotti}}]{Reimer2:14}%
  \BibitemOpen
  \bibfield  {author} {\bibinfo {author} {\bibfnamefont {C.}~\bibnamefont
  {Reimer}}, \bibinfo {author} {\bibfnamefont {L.}~\bibnamefont {Caspani}},
  \bibinfo {author} {\bibfnamefont {M.}~\bibnamefont {Clerici}}, \bibinfo
  {author} {\bibfnamefont {M.}~\bibnamefont {Ferrera}}, \bibinfo {author}
  {\bibfnamefont {M.}~\bibnamefont {Kues}}, \bibinfo {author} {\bibfnamefont
  {M.}~\bibnamefont {Peccianti}}, \bibinfo {author} {\bibfnamefont
  {A.}~\bibnamefont {Pasquazi}}, \bibinfo {author} {\bibfnamefont
  {L.}~\bibnamefont {Razzari}}, \bibinfo {author} {\bibfnamefont {B.~E.}\
  \bibnamefont {Little}}, \bibinfo {author} {\bibfnamefont {S.~T.}\
  \bibnamefont {Chu}}, \bibinfo {author} {\bibfnamefont {D.~J.}\ \bibnamefont
  {Moss}}, \ and\ \bibinfo {author} {\bibfnamefont {R.}~\bibnamefont
  {Morandotti}},\ }\href {\doibase 10.1364/OE.22.006535} {\bibfield  {journal}
  {\bibinfo  {journal} {Opt. Express}\ }\textbf {\bibinfo {volume} {22}},\
  \bibinfo {pages} {6535} (\bibinfo {year} {2014})}\BibitemShut {NoStop}%
\bibitem [{\citenamefont {Chiu}\ \emph {et~al.}(2014)\citenamefont {Chiu},
  \citenamefont {Chen}, \citenamefont {Chen}, \citenamefont {Yu}, \citenamefont
  {Chen},\ and\ \citenamefont {Chen}}]{PhysRevA.89.023839}%
  \BibitemOpen
  \bibfield  {author} {\bibinfo {author} {\bibfnamefont {C.-K.}\ \bibnamefont
  {Chiu}}, \bibinfo {author} {\bibfnamefont {Y.-H.}\ \bibnamefont {Chen}},
  \bibinfo {author} {\bibfnamefont {Y.-C.}\ \bibnamefont {Chen}}, \bibinfo
  {author} {\bibfnamefont {I.~A.}\ \bibnamefont {Yu}}, \bibinfo {author}
  {\bibfnamefont {Y.-C.}\ \bibnamefont {Chen}}, \ and\ \bibinfo {author}
  {\bibfnamefont {Y.-F.}\ \bibnamefont {Chen}},\ }\href {\doibase
  10.1103/PhysRevA.89.023839} {\bibfield  {journal} {\bibinfo  {journal} {Phys.
  Rev. A}\ }\textbf {\bibinfo {volume} {89}},\ \bibinfo {pages} {023839}
  (\bibinfo {year} {2014})}\BibitemShut {NoStop}%
\bibitem [{\citenamefont {Foster}\ \emph {et~al.}(2008)\citenamefont {Foster},
  \citenamefont {Turner}, \citenamefont {Lipson},\ and\ \citenamefont
  {Gaeta}}]{Foster:08}%
  \BibitemOpen
  \bibfield  {author} {\bibinfo {author} {\bibfnamefont {M.~A.}\ \bibnamefont
  {Foster}}, \bibinfo {author} {\bibfnamefont {A.~C.}\ \bibnamefont {Turner}},
  \bibinfo {author} {\bibfnamefont {M.}~\bibnamefont {Lipson}}, \ and\ \bibinfo
  {author} {\bibfnamefont {A.~L.}\ \bibnamefont {Gaeta}},\ }\href {\doibase
  10.1364/OE.16.001300} {\bibfield  {journal} {\bibinfo  {journal} {Opt.
  Express}\ }\textbf {\bibinfo {volume} {16}},\ \bibinfo {pages} {1300}
  (\bibinfo {year} {2008})}\BibitemShut {NoStop}%
\bibitem [{\citenamefont {Harada}\ \emph {et~al.}(2010)\citenamefont {Harada},
  \citenamefont {Takesue}, \citenamefont {Fukuda}, \citenamefont {Tsuchizawa},
  \citenamefont {Watanabe}, \citenamefont {Yamada}, \citenamefont {Tokura},\
  and\ \citenamefont {Itabashi}}]{Harada:2010aa}%
  \BibitemOpen
  \bibfield  {author} {\bibinfo {author} {\bibfnamefont {K.}~\bibnamefont
  {Harada}}, \bibinfo {author} {\bibfnamefont {H.}~\bibnamefont {Takesue}},
  \bibinfo {author} {\bibfnamefont {H.}~\bibnamefont {Fukuda}}, \bibinfo
  {author} {\bibfnamefont {T.}~\bibnamefont {Tsuchizawa}}, \bibinfo {author}
  {\bibfnamefont {T.}~\bibnamefont {Watanabe}}, \bibinfo {author}
  {\bibfnamefont {K.}~\bibnamefont {Yamada}}, \bibinfo {author} {\bibfnamefont
  {Y.}~\bibnamefont {Tokura}}, \ and\ \bibinfo {author} {\bibfnamefont
  {S.}~\bibnamefont {Itabashi}},\ }\href {\doibase 10.1109/JSTQE.2009.2023338}
  {\bibfield  {journal} {\bibinfo  {journal} {IEEE J. Sel. Top. Quantum
  Electron.}\ }\textbf {\bibinfo {volume} {16}},\ \bibinfo {pages} {325}
  (\bibinfo {year} {2010})}\BibitemShut {NoStop}%
\bibitem [{\citenamefont {Eggleton}, \citenamefont {Luther-Davies},\ and\
  \citenamefont {Richardson}(2011)}]{Eggleton2011Chalcoge}%
  \BibitemOpen
  \bibfield  {author} {\bibinfo {author} {\bibfnamefont {B.~J.}\ \bibnamefont
  {Eggleton}}, \bibinfo {author} {\bibfnamefont {B.}~\bibnamefont
  {Luther-Davies}}, \ and\ \bibinfo {author} {\bibfnamefont {K.}~\bibnamefont
  {Richardson}},\ }\href {http://dx.doi.org/10.1038/nphoton.2011.309}
  {\bibfield  {journal} {\bibinfo  {journal} {Nature Photon.}\ }\textbf
  {\bibinfo {volume} {5}},\ \bibinfo {pages} {141} (\bibinfo {year}
  {2011})}\BibitemShut {NoStop}%
\bibitem [{\citenamefont {Dot}\ \emph {et~al.}(2014)\citenamefont {Dot},
  \citenamefont {Meyer-Scott}, \citenamefont {Ahmad}, \citenamefont
  {Rochette},\ and\ \citenamefont {Jennewein}}]{PhysRevA.90.043808}%
  \BibitemOpen
  \bibfield  {author} {\bibinfo {author} {\bibfnamefont {A.}~\bibnamefont
  {Dot}}, \bibinfo {author} {\bibfnamefont {E.}~\bibnamefont {Meyer-Scott}},
  \bibinfo {author} {\bibfnamefont {R.}~\bibnamefont {Ahmad}}, \bibinfo
  {author} {\bibfnamefont {M.}~\bibnamefont {Rochette}}, \ and\ \bibinfo
  {author} {\bibfnamefont {T.}~\bibnamefont {Jennewein}},\ }\href {\doibase
  10.1103/PhysRevA.90.043808} {\bibfield  {journal} {\bibinfo  {journal} {Phys.
  Rev. A}\ }\textbf {\bibinfo {volume} {90}},\ \bibinfo {pages} {043808}
  (\bibinfo {year} {2014})}\BibitemShut {NoStop}%
\bibitem [{\citenamefont {Slusher}\ \emph {et~al.}(2004)\citenamefont
  {Slusher}, \citenamefont {Lenz}, \citenamefont {Hodelin}, \citenamefont
  {Sanghera}, \citenamefont {Shaw},\ and\ \citenamefont
  {Aggarwal}}]{Slusher:04}%
  \BibitemOpen
  \bibfield  {author} {\bibinfo {author} {\bibfnamefont {R.~E.}\ \bibnamefont
  {Slusher}}, \bibinfo {author} {\bibfnamefont {G.}~\bibnamefont {Lenz}},
  \bibinfo {author} {\bibfnamefont {J.}~\bibnamefont {Hodelin}}, \bibinfo
  {author} {\bibfnamefont {J.}~\bibnamefont {Sanghera}}, \bibinfo {author}
  {\bibfnamefont {L.~B.}\ \bibnamefont {Shaw}}, \ and\ \bibinfo {author}
  {\bibfnamefont {I.~D.}\ \bibnamefont {Aggarwal}},\ }\href@noop {} {\bibfield
  {journal} {\bibinfo  {journal} {J. Opt. Soc. Am. B}\ }\textbf {\bibinfo
  {volume} {21}},\ \bibinfo {pages} {1146} (\bibinfo {year}
  {2004})}\BibitemShut {NoStop}%
\bibitem [{\citenamefont {Ahmad}\ and\ \citenamefont
  {Rochette}(2012)}]{Ahmad:12}%
  \BibitemOpen
  \bibfield  {author} {\bibinfo {author} {\bibfnamefont {R.}~\bibnamefont
  {Ahmad}}\ and\ \bibinfo {author} {\bibfnamefont {M.}~\bibnamefont
  {Rochette}},\ }\href@noop {} {\bibfield  {journal} {\bibinfo  {journal} {Opt.
  Express}\ }\textbf {\bibinfo {volume} {20}},\ \bibinfo {pages} {9572}
  (\bibinfo {year} {2012})}\BibitemShut {NoStop}%
\bibitem [{\citenamefont {Baker}\ and\ \citenamefont
  {Rochette}(2012)}]{Baker2012High-Non}%
  \BibitemOpen
  \bibfield  {author} {\bibinfo {author} {\bibfnamefont {C.}~\bibnamefont
  {Baker}}\ and\ \bibinfo {author} {\bibfnamefont {M.}~\bibnamefont
  {Rochette}},\ }\bibfield  {booktitle} {\emph {\bibinfo {booktitle} {Photonics
  Journal, IEEE}},\ }\href {\doibase 10.1109/JPHOT.2012.2202103} {\bibfield
  {journal} {\bibinfo  {journal} {Photonics Journal, IEEE}\ }\textbf {\bibinfo
  {volume} {4}},\ \bibinfo {pages} {960} (\bibinfo {year} {2012})}\BibitemShut {NoStop}%
\bibitem [{\citenamefont {Agrawal}(2013)}]{tagkey2013i}%
  \BibitemOpen
  \bibfield  {author} {\bibinfo {author} {\bibfnamefont {G.~P.}\ \bibnamefont
  {Agrawal}},\ }\href {\doibase
  http://dx.doi.org/10.1016/B978-0-12-397023-7.00018-8} {\emph {\bibinfo
  {title} {Nonlinear Fiber Optics}}},\ \bibinfo {edition} {5th}\ ed.\ (\bibinfo
   {publisher} {Academic Press},\ \bibinfo {address} {Boston},\ \bibinfo {year}
  {2013})\BibitemShut {NoStop}%
\bibitem [{\citenamefont {Baker}\ and\ \citenamefont
  {Rochette}(2010)}]{Baker:10}%
  \BibitemOpen
  \bibfield  {author} {\bibinfo {author} {\bibfnamefont {C.}~\bibnamefont
  {Baker}}\ and\ \bibinfo {author} {\bibfnamefont {M.}~\bibnamefont
  {Rochette}},\ }\href@noop {} {\bibfield  {journal} {\bibinfo  {journal} {Opt.
  Express}\ }\textbf {\bibinfo {volume} {18}},\ \bibinfo {pages} {12391}
  (\bibinfo {year} {2010})}\BibitemShut {NoStop}%
\bibitem [{\citenamefont {Kasarova}\ \emph {et~al.}(2007)\citenamefont
  {Kasarova}, \citenamefont {Sultanova}, \citenamefont {Ivanov},\ and\
  \citenamefont {Nikolov}}]{Kasarova20071481}%
  \BibitemOpen
  \bibfield  {author} {\bibinfo {author} {\bibfnamefont {S.~N.}\ \bibnamefont
  {Kasarova}}, \bibinfo {author} {\bibfnamefont {N.~G.}\ \bibnamefont
  {Sultanova}}, \bibinfo {author} {\bibfnamefont {C.~D.}\ \bibnamefont
  {Ivanov}}, \ and\ \bibinfo {author} {\bibfnamefont {I.~D.}\ \bibnamefont
  {Nikolov}},\ }\href {\doibase http://dx.doi.org/10.1016/j.optmat.2006.07.010}
  {\bibfield  {journal} {\bibinfo  {journal} {Opt. Mater.}\ }\textbf {\bibinfo
  {volume} {29}},\ \bibinfo {pages} {1481 } (\bibinfo {year}
  {2007})}\BibitemShut {NoStop}%
\bibitem [{\citenamefont {Yan}\ \emph {et~al.}(2012)\citenamefont {Yan},
  \citenamefont {Hamel}, \citenamefont {Heinrichs}, \citenamefont {Jiang},
  \citenamefont {Itzler},\ and\ \citenamefont
  {Jennewein}}]{:/content/aip/journal/rsi/83/7/10.1063/1.4732813}%
  \BibitemOpen
  \bibfield  {author} {\bibinfo {author} {\bibfnamefont {Z.}~\bibnamefont
  {Yan}}, \bibinfo {author} {\bibfnamefont {D.~R.}\ \bibnamefont {Hamel}},
  \bibinfo {author} {\bibfnamefont {A.~K.}\ \bibnamefont {Heinrichs}}, \bibinfo
  {author} {\bibfnamefont {X.}~\bibnamefont {Jiang}}, \bibinfo {author}
  {\bibfnamefont {M.~A.}\ \bibnamefont {Itzler}}, \ and\ \bibinfo {author}
  {\bibfnamefont {T.}~\bibnamefont {Jennewein}},\ }\href@noop {} {\bibfield
  {journal} {\bibinfo  {journal} {Rev. Sci. Instrum.}\ }\textbf {\bibinfo
  {volume} {83}},\ \bibinfo {pages} {073105} (\bibinfo {year}
  {2012})}\BibitemShut {NoStop}%
\bibitem [{\citenamefont {Steinlechner}\ \emph {et~al.}(2013)\citenamefont
  {Steinlechner}, \citenamefont {Ramelow}, \citenamefont {Jofre}, \citenamefont
  {Gilaberte}, \citenamefont {Jennewein}, \citenamefont {Torres}, \citenamefont
  {Mitchell},\ and\ \citenamefont {Pruneri}}]{Steinlechner:13}%
  \BibitemOpen
  \bibfield  {author} {\bibinfo {author} {\bibfnamefont {F.}~\bibnamefont
  {Steinlechner}}, \bibinfo {author} {\bibfnamefont {S.}~\bibnamefont
  {Ramelow}}, \bibinfo {author} {\bibfnamefont {M.}~\bibnamefont {Jofre}},
  \bibinfo {author} {\bibfnamefont {M.}~\bibnamefont {Gilaberte}}, \bibinfo
  {author} {\bibfnamefont {T.}~\bibnamefont {Jennewein}}, \bibinfo {author}
  {\bibfnamefont {J.~P.}\ \bibnamefont {Torres}}, \bibinfo {author}
  {\bibfnamefont {M.~W.}\ \bibnamefont {Mitchell}}, \ and\ \bibinfo {author}
  {\bibfnamefont {V.}~\bibnamefont {Pruneri}},\ }\href {\doibase
  10.1364/OE.21.011943} {\bibfield  {journal} {\bibinfo  {journal} {Opt.
  Express}\ }\textbf {\bibinfo {volume} {21}},\ \bibinfo {pages} {11943}
  (\bibinfo {year} {2013})}\BibitemShut {NoStop}%
\bibitem [{\citenamefont {Engin}\ \emph {et~al.}(2013)\citenamefont {Engin},
  \citenamefont {Bonneau}, \citenamefont {Natarajan}, \citenamefont {Clark},
  \citenamefont {Tanner}, \citenamefont {Hadfield}, \citenamefont {Dorenbos},
  \citenamefont {Zwiller}, \citenamefont {Ohira}, \citenamefont {Suzuki},
  \citenamefont {Yoshida}, \citenamefont {Iizuka}, \citenamefont {Ezaki},
  \citenamefont {O'Brien},\ and\ \citenamefont {Thompson}}]{Engin:13}%
  \BibitemOpen
  \bibfield  {author} {\bibinfo {author} {\bibfnamefont {E.}~\bibnamefont
  {Engin}}, \bibinfo {author} {\bibfnamefont {D.}~\bibnamefont {Bonneau}},
  \bibinfo {author} {\bibfnamefont {C.~M.}\ \bibnamefont {Natarajan}}, \bibinfo
  {author} {\bibfnamefont {A.~S.}\ \bibnamefont {Clark}}, \bibinfo {author}
  {\bibfnamefont {M.~G.}\ \bibnamefont {Tanner}}, \bibinfo {author}
  {\bibfnamefont {R.~H.}\ \bibnamefont {Hadfield}}, \bibinfo {author}
  {\bibfnamefont {S.~N.}\ \bibnamefont {Dorenbos}}, \bibinfo {author}
  {\bibfnamefont {V.}~\bibnamefont {Zwiller}}, \bibinfo {author} {\bibfnamefont
  {K.}~\bibnamefont {Ohira}}, \bibinfo {author} {\bibfnamefont
  {N.}~\bibnamefont {Suzuki}}, \bibinfo {author} {\bibfnamefont
  {H.}~\bibnamefont {Yoshida}}, \bibinfo {author} {\bibfnamefont
  {N.}~\bibnamefont {Iizuka}}, \bibinfo {author} {\bibfnamefont
  {M.}~\bibnamefont {Ezaki}}, \bibinfo {author} {\bibfnamefont {J.~L.}\
  \bibnamefont {O'Brien}}, \ and\ \bibinfo {author} {\bibfnamefont {M.~G.}\
  \bibnamefont {Thompson}},\ }\href {\doibase 10.1364/OE.21.027826} {\bibfield
  {journal} {\bibinfo  {journal} {Opt. Express}\ }\textbf {\bibinfo {volume}
  {21}},\ \bibinfo {pages} {27826} (\bibinfo {year} {2013})}\BibitemShut
  {NoStop}%
\bibitem [{\citenamefont {He}\ \emph {et~al.}(2012)\citenamefont {He},
  \citenamefont {Xiong}, \citenamefont {Clark}, \citenamefont {Collins},
  \citenamefont {Gai}, \citenamefont {Choi}, \citenamefont {Madden},
  \citenamefont {Luther-Davies},\ and\ \citenamefont
  {Eggleton}}]{:/content/aip/journal/jap/112/12/10.1063/1.4769740}%
  \BibitemOpen
  \bibfield  {author} {\bibinfo {author} {\bibfnamefont {J.}~\bibnamefont
  {He}}, \bibinfo {author} {\bibfnamefont {C.}~\bibnamefont {Xiong}}, \bibinfo
  {author} {\bibfnamefont {A.~S.}\ \bibnamefont {Clark}}, \bibinfo {author}
  {\bibfnamefont {M.~J.}\ \bibnamefont {Collins}}, \bibinfo {author}
  {\bibfnamefont {X.}~\bibnamefont {Gai}}, \bibinfo {author} {\bibfnamefont
  {D.-Y.}\ \bibnamefont {Choi}}, \bibinfo {author} {\bibfnamefont {S.~J.}\
  \bibnamefont {Madden}}, \bibinfo {author} {\bibfnamefont {B.}~\bibnamefont
  {Luther-Davies}}, \ and\ \bibinfo {author} {\bibfnamefont {B.~J.}\
  \bibnamefont {Eggleton}},\ }\href {\doibase
  http://dx.doi.org/10.1063/1.4769740} {\bibfield  {journal} {\bibinfo
  {journal} {J. Appl. Phys.}\ }\textbf {\bibinfo {volume} {112}},\ \bibinfo
  {eid} {123101} (\bibinfo {year} {2012})}\BibitemShut {NoStop}%
\bibitem [{\citenamefont {Collins}\ \emph {et~al.}(2012)\citenamefont
  {Collins}, \citenamefont {Clark}, \citenamefont {He}, \citenamefont {Choi},
  \citenamefont {Williams}, \citenamefont {Judge}, \citenamefont {Madden},
  \citenamefont {Withford}, \citenamefont {Steel}, \citenamefont
  {Luther-Davies}, \citenamefont {Xiong},\ and\ \citenamefont
  {Eggleton}}]{Collins:12}%
  \BibitemOpen
  \bibfield  {author} {\bibinfo {author} {\bibfnamefont {M.~J.}\ \bibnamefont
  {Collins}}, \bibinfo {author} {\bibfnamefont {A.~S.}\ \bibnamefont {Clark}},
  \bibinfo {author} {\bibfnamefont {J.}~\bibnamefont {He}}, \bibinfo {author}
  {\bibfnamefont {D.-Y.}\ \bibnamefont {Choi}}, \bibinfo {author}
  {\bibfnamefont {R.~J.}\ \bibnamefont {Williams}}, \bibinfo {author}
  {\bibfnamefont {A.~C.}\ \bibnamefont {Judge}}, \bibinfo {author}
  {\bibfnamefont {S.~J.}\ \bibnamefont {Madden}}, \bibinfo {author}
  {\bibfnamefont {M.~J.}\ \bibnamefont {Withford}}, \bibinfo {author}
  {\bibfnamefont {M.~J.}\ \bibnamefont {Steel}}, \bibinfo {author}
  {\bibfnamefont {B.}~\bibnamefont {Luther-Davies}}, \bibinfo {author}
  {\bibfnamefont {C.}~\bibnamefont {Xiong}}, \ and\ \bibinfo {author}
  {\bibfnamefont {B.~J.}\ \bibnamefont {Eggleton}},\ }\href {\doibase
  10.1364/OL.37.003393} {\bibfield  {journal} {\bibinfo  {journal} {Opt.
  Lett.}\ }\textbf {\bibinfo {volume} {37}},\ \bibinfo {pages} {3393} (\bibinfo
  {year} {2012})}\BibitemShut {NoStop}%
\end{thebibliography}
\end{document}